\documentstyle[12pt]{article}        
\newlength{\dinwidth}                       
\newlength{\dinmargin}                      
\def\lsim{\mathrel{\rlap{\lower4pt\hbox{\hskip1pt$\sim$}}
    \raise1pt\hbox{$<$}}}                
\def\gsim{\mathrel{\rlap{\lower4pt\hbox{\hskip1pt$\sim$}}
    \raise1pt\hbox{$>$}}}                
\begin{document}
\vspace*{1cm}
\begin{center}  \begin{Large} \begin{bf}
1995-96 HERA WORKSHOP \\
BEYOND THE STANDARD MODEL GROUP SUMMARY \\
  \end{bf}  \end{Large}
  \vspace*{5mm}
  \begin{large}
H. Dreiner$^a$, H.-U. Martyn$^{b}$, S. Ritz$^c$, D. Wyler$^d$\\ 
  \end{large}
\end{center}
$^a$ {\em Rutherford Appleton Laboratory, Chilton, Didcot, Oxon, UK}\\
$^b$ {\em I. Physikalisches Institute, RWTH Aachen, Germany}\\
$^c$ {\em Columbia University, New York, N.Y., U.S.A.}\\
$^d$ {\em Universit\"at Z\"urich, Z\"urich, Switzerland}
\begin{quotation}
\noindent
{\bf Abstract:}
We summarize the work done in the ``Beyond the Standard Model'' group
of the 1995-96 HERA Workshop.  Of the various proposed HERA upgrades, the 
luminosity improvement is the most important for this physics.  With an 
integrated luminosity of 1 $fb^{-1}$, collected by 2005, HERA will remain 
a competitive and potentially fruitful facility for new physics searches.
\end{quotation}
\section{Summary Statement}

The goal of the 1995-96 HERA Workshop was to work out the implications 
for  physics 
with the proposed upgrades to HERA.  Since, by construction, the
Beyond the Standard Model group must study highly speculative topics
(most, if not all, known extensions to the Standard Model are certainly
wrong), we studied whether potentially interesting physics could be made
more accessible by the upgrades.

Of the proposed upgrades, the luminosity enhancement is {\em a priori} the most
important for this physics.   Without substantially enlarging the data set
in relatively short periods of time, exotic physics searches stall.  In
addition, the competition from other facilities, in particular LEP and the
Tevatron, is stiff.   The results of our studies indicate that a
luminosity upgrade is essential if the HERA program is to remain
interesting and competitive in this area.

Since we are going beyond the Standard Model, the topics we studied may
be organized by {\em how} far beyond the Standard Model they lie, as
follows. The individual contributions can be found in full at:
http://ucosun.desy.de/~heraws96/proceedings/beyondSM/

\subsection{Higgses}
We start by revisiting searches for Higgs bosons.  Standard Model Higgs
physics has been deemed hopeless at HERA, mainly due to the small
production cross-section\cite{ref-workshop91}.  (With an integrated 
luminosity of 1 $fb^{-1}$,
HERA might actually produce a few Standard Model Higgs bosons in
currently allowed mass ranges, but isolating these few events from the
background still looks hopeless.)  Plausible non-standard Higgs sectors
offer more possibilities.  Many models, including the minimal SUSY
extension to the Standard Model, include two Higgs doublets, generating
five physical particle degrees of freedom: two neutral scalars ($H^0$ 
and $h^0$), a
neutral pseudoscalar ($A$), and two charged scalars ($H^\pm$);  and 
introducing two
parameters that modify the couplings: a mixing angle for the neutral
scalars ($\alpha$) and the ratio of the vacuum expectation values for the two
doublets ($\tan\beta$).  In Minimal SUSY  these parameters are constrained 
to regions that keep these light Higgses out of HERA's reach, when taking
the LEP bounds into account. However, there is 
good reason to look anyway.  The LEP program is steadily eating away the
remaining allowed regions and, SUSY aside, non-minimal Higgs sectors have been 
proposed as mechanisms for a wide variety of phenomena, {\em e.g.},   
electroweak CP violation, the suppression of strong CP violation, and
neutrino mass generation.  Such a search is well-motivated.  

Maria Krawczyk has studied the phenomenology of a general two-doublet model.
She found that there are regions of ($\alpha ,\beta$) that are not ruled out
by LEP, even for very light Higgs massless of a few GeV.  Such Higgses
{\em can} be produced at HERA via photoproduction.  The resolved 
process, $gg\rightarrow h$, results in a $b\overline{b}$ final state (or
a $\tau^+ \tau^-$ final state for very light $h$), while
the direct process $\gamma g \rightarrow b\overline{b}h$ results in an
enticing four-$b$ final state (or a $b\overline{b}\tau^+ \tau^-$ final state
for very light $h$).  The success of these searches will depend on how well
the signals can be isolated from the backgrounds.  Krawczyk and Ritz are 
producing\footnote{Consult the web page for details.} generators to study 
these processes.  There may, after all, be a Higgs for HERA, in photoproduction.

\subsection{Contact Interactions and Compositeness}

Moving somewhat further beyond the Standard Model, the study of contact
interactions provides a model-independent way to parametrize the sensitivity to 
new physics.  Jason Gilmore has studied the sensitivity to $eeqq$ contact
interactions, as well as to finite quark radii.  He concludes that integrated 
luminosities of order 200-500 $pb^{-1}$ in 
both $e^-p$ and $e^+p$ are necessary to probe distances shorter 
than $10^{-16} cm$ and contact interaction scales comparable to those accessible
at the Tevatron.

\subsection{Lepton Flavor Violation}

With the relatively clean HERA environment, lepton flavor violating process
can be probed in a straightforward and general manner:  a high $Q^2$
DIS-like final state is sought with a $\mu$ or $\tau$ replacing the scattered
lepton beam particle.  There are already results from ZEUS\cite{ref-zeuslfv}
and H1\cite{ref-h1lfv}.  Frank Sciulli and Songhoon Yang have extended their
original analysis of leptoquarks with 2nd and 3rd generation couplings
to the 1 $fb^{-1}$ case, and show that HERA will have the world's best
sensitivity to many types of these particles.

\subsection{Heavy Neutral Leptons}

Frank Sciulli and Larry Wai have studied an interesting case for HERA: a neutral 
right handed lepton with moderate mass may have escaped detection {\em if}
it is more massive than the associated right-handed $W_R$ boson.  In that case, the
$W_R$ will decay only to a pair of jets with no missing momentum, and the existing 
experimental limits are not valid for $W_R$ masses {\em below} 100 GeV.  Again, 
high luminosities allow a significant discovery potential.

\subsection{Supersymmetry}
Supersymmetry is the most widely studied extension of the Standard
Model.  Members of our group have focused on three distinct models of 
supersymmetry:
(a) the minimal supersymmetric standard model (MSSM) with conserved
R-parity and a gluino mass above the Tevatron bound of 140 GeV;
(b) R-parity violation through the $LQ{\bar D}$\cite{ref-whatthehellisLQD?} 
operator, and (c) the light gluino scenario where $M_{gluino}\lsim 1 GeV$.

\subsubsection{$R_p$ Conserving}
In the MSSM we considered the two processes
\begin{eqnarray}
e^-+q & \rightarrow &{\tilde e}^- + {\tilde q} \\
e^-+q&\rightarrow & {\tilde e}^-+{\tilde\chi}^0_1+q
\end{eqnarray}
where ${\tilde\chi}^0_1$ is the lightest neutralino. 

For the first
process, studied by Peter Schleper, the present
bounds from LEP1.3 are comparable to those from HERA\cite{ref-HERAschleper}. 
The overall sensitivity with 500 pb$^{-1}$ 
is comparable to LEP2. Therefore the sooner we obtain the upgrade 
the more likely HERA can remain in competition. Optimistically, if LEP2
discovers this process then HERA can also access this physics.  
Pessimistically, if one applies the
model-dependent scalar quark bounds from the Tevatron then HERA is 
not competitive.  The second process was investigated by Massimo Corradi.  
Unfortunately, LEP1.3 has already excluded the region of parameter space 
to which HERA can ever be sensitive\cite{ref-LEP1.3bounds} in this particular
model.

\subsubsection{$R_p$ Violating}
R-parity violation has several Yukawa couplings beyond those of the MSSM
which violate either lepton number or baryon number.  HERA is particularly
sensitive to a subset of the lepton number-violating couplings, denoted
$LQ{\bar D}$\cite{ref-whatthehellisLQD?} since they lead to resonant
scalar quark production:
\begin{eqnarray}
e^-+q & \rightarrow & {\tilde q}\, \rightarrow\, q'+{\tilde\chi}
\end{eqnarray}
where ${\tilde\chi}$ represents a general gaugino (neutralino or chargino).
This has been studied in much greater detail by Dreiner, Perez, and Sirois,
extending the scalar squark decays to the entire supersymmetric spectrum.
HERA remains the best machine for this process.

Previously, Dreiner and Morawitz studied the process\cite{ref-herbipeter}
\begin{eqnarray}
e^-+q&\rightarrow & {\tilde e}^- + {\tilde q}, \quad ({\tilde q},
{\tilde e})\rightarrow (q, e) + {\tilde\chi}^0_1.
\end{eqnarray}
HERA is sensitive to the case when the ${\tilde\chi}^0_1$ decays via lepton 
number-violating
operators $L_3Q{\bar D}$ or $L_3Q{\bar E}$.  These lead to tau-lepton final
states.  HERA is the best machine to test these operators.

\subsubsection{Light Gluino}

The exclusion of a light gluino with mass below 5-10 GeV is debated 
within the supersymmetry community. If a light gluino exists it should be
copiously produced in photoproduction at HERA\cite{ref-lightgluinosigma}, 
and will likely
hadronize as a long-lived, electrically neutral
particle\cite{ref-Glennys}.  Marc David has studied the possibilities of
detecting such processes using topologies of energy deposits in the 
H1 calorimeter. 

For multi-jet processes in DIS involving a light gluino, 
Graudenz {\em et al.} confirm that 
extracting a
signal from the very large QCD background would require more clever
analyses than jet-angle variables alone. 

It is worth noting that a recent 
reanalysis\cite{ref-OPALgluino} of OPAL data may have closed the light 
gluino window definitively.

\subsection{Other New Particles}

With general scaling rules, Uli Martyn has extended previous workshop results
and existing HERA results to the 1 $fb^{-1}$ domain.  Topics include exotic
leptons, excited leptons, excited quarks, leptoquarks, leptogluons, new vector
bosons, compositeness, and quark form factors.  With an integrated luminosity
of 1 $fb^{-1}$, HERA will remain an excellent facility
to search for most of these extensions to the Standard Model.

We also have a contribution from Bl\"{u}mlein {\em et al.} explicitly 
considering  scalar and vector leptoquark pair production.  The advantage 
here is that
leptoquark couplings to gauge particles are determined, in contrast to the 
more familiar single leptoquark production case where the Yukawa coupling is
a free parameter.  In this mode, HERA is likely to be
most competitive, if at all, in
searches for leptoquark pairs that decay to 3rd generation states ($b, \tau$).

\section{Conclusions}

With a substantial luminosity upgrade HERA will continue to have good 
discovery potential for most of the topics we have studied.  In addition to
the amount of integrated luminosity, {\em when} that luminosity is delivered 
also matters.  Table \ref{tab-lumiprofile} shows luminosity profiles in 
different scenarios.  The numbers represent only our best guesses, but they
illustrate a point: as the annual integrated luminosity asymptotes, the time
required to acquire substantially increased statistics grows.  One figure
of merit is the time to double the existing data sample.  If HERA asymptotes
to 35 $pb^{-1}/y$, it will be difficult to wait to accumulate even 
250 $pb^{-1}$.
By contrast, if the asymptotic value is 170-200 $pb^{-1}/y$, integrated delivered 
luminosities in the neighborhood of 1 $fb^{-1}$ can be accumulated in a timely
manner, with substantial new data sets each year to maintain an exciting 
program.  Finally, since plans for the Tevatron indicate an integrated 
luminosity of 33 $fb^{-1}$ by that same time, HERA must certainly upgrade to 
be of interest for physics beyond the Standard Model.

\begin{table}
\begin{displaymath}
\begin{tabular}{|r||r|r|r||r|r|r|}
\hline
Year & Annual-35 & Int & 2yr ratio & Annual-170 & Int & 2yr ratio \\
\hline\hline
1993 & 1& 1& & 1& 1&\\
\hline
1994 & 6& 7& & 6& 7&\\
\hline
1995 & 12& 19&19 &12 &19 &19\\
\hline
1996 & 15&34 &4.9 &15 &34 &4.9\\
\hline
1997 & 30&64 &3.4 &30 &64 &3.4\\
\hline
1998 & 35&99 &2.9 &50 &114 &3.4\\
\hline
1999 & 35&134 &2.1 & 100& 214&3.3\\
\hline
2000 &35 &\large{\bf 169} &1.7 &125 &339 &3.0\\
\hline
2001 &35 &204 &1.5 &150 &489 &2.3\\
\hline
2002 &35 &239 &1.4 &170 &659 &1.9\\
\hline
2003 &35 &274 &1.3 &170 &\large{\bf 829} &1.7\\
\hline
2004 &35 &309 &1.3 &170 &999 &1.5\\
\hline
2005 &35 &344 &1.3 &170 &1169 &1.4\\
\hline
\end{tabular}
\end{displaymath}
\caption{{\em Two versions of the future.  Luminosity profiles for a
machine that asymptotes to 35 $pb^{-1}/y$ (first set of columns) and
170 $pb^{-1}/y$ (second set of columns).  In each group of columns, the first
gives the delivered luminosity that year, the second gives the running
total collected since turn-on, and the third gives the ratio of the running
total of that year to that amount two years earlier.  When this ratio falls
below 2, we have failed to double our statistics within two years (see text).
The running total at that point is given by the bold numbers for the two
scenarios.}}
\label{tab-lumiprofile}
\end{table}

\end{document}